\documentclass[journal,twoside,web]{IEEEtran}
\usepackage{generic}
\usepackage{amsmath,amssymb,amsfonts}
\usepackage{graphicx}
\usepackage{textcomp,nicefrac}

\usepackage{cite}
\usepackage{amsmath,amssymb,amsfonts}
\usepackage{algorithmic}
\usepackage{graphicx}
\usepackage{textcomp}

\usepackage{url} 

\usepackage[colorlinks=true, citecolor=blue,linkcolor=blue,urlcolor=blue]{hyperref} 

\newcommand*{\figureref}[2][]{%
  \hyperref[{fig:#2}]{%
    Figure~\ref*{fig:#2}%
    \ifx\\#1\\%
    \else
      \,#1%
    \fi
  }%
}
\newcommand*{\figref}[2][]{%
  \hyperref[{fig:#2}]{%
    Fig.~\ref*{fig:#2}%
    \ifx\\#1\\%
    \else
      \,#1%
    \fi
  }%
}

\newcommand*{\tabref}[2][]{%
  \hyperref[{tab:#2}]{%
    Table~\ref*{tab:#2}%
    \ifx\\#1\\%
    \else
      \,#1%
    \fi
  }%
}

\usepackage{xcolor}
\usepackage{pict2e}

\newsavebox{\ORCIDlogo}
\savebox{\ORCIDlogo}{%
\setlength{\unitlength}{\dimexpr 1em/256\relax}%
\begin{picture}(256,256)%
  \color[HTML]{A6CE39}\put(128,128){\circle*{256}}%
  \color{white}%
  \put(78.6,199.2){\circle*{20}}%
  \moveto(70.9,176,9)\lineto(86.3,176,9)\lineto(86.3,69.8)\lineto(70.9,69.8)%
  \closepath\fillpath%
  \moveto(108.9,176.9)\lineto(150.5,176.9)%
  \curveto(190.1,176.9)(207.5,148.6)(207.5 ,123.3)%
  \curveto(207.5,95,8)(186,69.7)(150.7,69.7)%
  \lineto(108.9,69.7)%
  \closepath\fillpath%
  \color[HTML]{A6CE39}%
  \moveto(124.3,83.6)\lineto(148.8,83.6)%
  \curveto(183.7,83.6)(191.7,110.1)(191.7,123.3)%
  \curveto(191.7,144.8)(178,163)(148,163)%
  \lineto(124.3,163)%
  \closepath\fillpath%
\end{picture}%
}

\newcommand\orcidicon[1]{\href{https://orcid.org/#1}{\usebox{\ORCIDlogo}}}

\usepackage{multirow}
\usepackage{booktabs}

\usepackage{float}
\usepackage{makecell}
\usepackage{boondox-calo}
\usepackage{lipsum}
\usepackage{longtable}
\usepackage[flushleft]{threeparttable}

\def\BibTeX{{\rm B\kern-.05em{\sc i\kern-.025em b}\kern-.08em
    T\kern-.1667em\lower.7ex\hbox{E}\kern-.125emX}}

\begin{document}
\title{On the Vulnerability of UMOSFETs in Terrestrial Radiation Environments}

\author{Saulo G. Alberton\orcidicon{0000-0002-7390-3660},
Alexis C. V. Bôas\orcidicon{0000-0002-5975-5518},
Jeffery Wyss\orcidicon{0000-0002-8277-4012}, 
Vitor A. P. Aguiar\orcidicon{0000-0001-6199-0800}, 
Matheus S. Pereira\orcidicon{0000-0002-4141-7166},
Luca Silvestrin\orcidicon{0000-0003-0551-242X},
Serena Mattiazzo\orcidicon{0000-0001-8255-3474},
Alessandro Paccagnella\orcidicon{0000-0002-6850-4286}, \textit{Senior Member, IEEE}, 
Carlo Cazzaniga\orcidicon{0000-0002-3110-0253},
Maria Kastriotou,
Christopher Frost\orcidicon{0000-0003-3541-6527}, and 
Nilberto H. Medina\orcidicon{0000-0003-0650-6507}, \textit{Member, IEEE}
\thanks{This study was financed in part by the Coordenação de Aperfeiçoamento de Pessoal de Nível Superior - Brasil (CAPES) - Financial Code 001; CNPq (Nos. 404054/2023-4 and 422360/2023-6); FAPESP (No. 2023/16053-8); FINEP (No. 01.12.0224.00); INCT-FNA (No. 464898/2014-5); and RADNEXT project (No. 101008126).}
\thanks{S. G. Alberton is with the Instituto de Fisica, Universidade de Sao Paulo, São Paulo, SP, Brazil, and also with the Dipartimento di Ingegneria dell’Informazione, Università degli Studi di Padova, Padova 35131, Italy (e-mails: alberton@if.usp.br, albertonsg@dei.unipd.it).}
\thanks{A. C. V. Bôas is with the Centro Universitário FEI, São Bernardo do Campo, SP, Brazil.}
\thanks{J. Wyss is with the Dipartimento di Ingegneria Civile e Meccanica, Università degli Studi di Cassino e del Lazio Meridionale, Cassino 03043, Italy, and also with the Istituto Nazionale di Fisica Nucleare, Padova 35131, Italy.}
\thanks{V. A. P. Aguiar, M. S. Pereira, and N. H. Medina are with the Instituto de Fisica, Universidade de Sao Paulo, São Paulo, SP, Brazil.}
\thanks{L. Silvestrin and S. Mattiazzo are with the Dipartimento di Fisica e Astronomia ``Galileo Galilei'', Università degli Studi di Padova, Padova 35131, Italy, and also with the Istituto Nazionale di Fisica Nucleare - Sezione di Padova, Padova 35131, Italy.}
\thanks{A. Paccagnella is with the Dipartimento di Ingegneria dell’Informazione, Università degli Studi di Padova, Padova 35131, Italy.}
\thanks{C. Cazzaniga, M. Kastriotou, and C. Frost are with the ISIS Neutron and Muon Source, Rutherford Appleton Laboratory, OX11 0QX, Didcot, UK.}
}

\maketitle

\markboth{}{}

\begin{abstract}
The vulnerability of prominent silicon-based U-shaped Metal-Oxide-Semiconductor Field Effect Transistors (UMOSFET) to destructive radiation effects when operating in terrestrial atmospheric environments is addressed. It is known that secondary particles from nuclear reactions between atmospheric neutrons and the constituent materials of electronic devices can trigger Single-Event Burnout (SEB) destructive failure in power MOSFETs. 
The susceptibility of UMOSFETs to SEBs induced by atmospheric neutrons in accelerated tests are compared to that of similarly rated traditional Double-diffused MOSFET (DMOSFET) counterparts.
Computational simulations are conducted to elucidate failure mechanisms and propose strategies to potentially enhance the survivability of next-generation UMOSFETs in high-reliability power systems operating on Earth.
\end{abstract}

\begin{IEEEkeywords}
Radiation effects, power transistor, UMOSFET, trench MOSFET, DMOSFET, atmospheric neutrons, single-event effect, single-event burnout, single-event gate rupture.
\end{IEEEkeywords}

\section{Introduction}
\label{sec:introduction}
\begin{figure}[t]   
    \centering
    \includegraphics[width=0.43\textwidth]{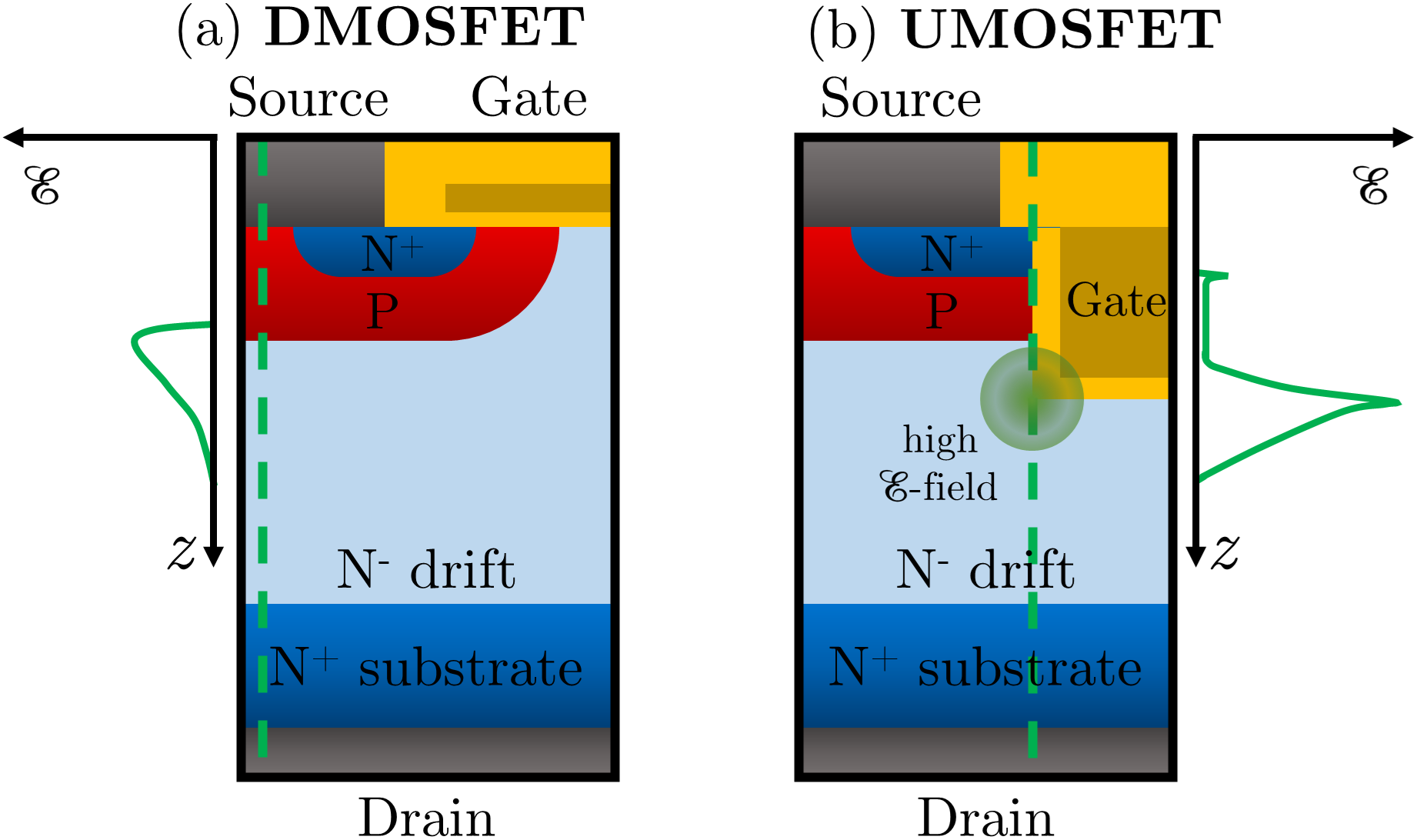}
    \caption[]{Half-cell cross sections of (a) planar DMOS and (b) UMOS power FETs. Electric field distributions at the positions marked by the dashed green lines are also depicted.}
    \label{fig:umos-dmos}
    \vspace{-5mm}
\end{figure}

\IEEEPARstart{T}{he} U-shaped Metal-Oxide-Semiconductor Field Effect Transistor (UMOS or trench FET) is currently one of the most popular semiconductor power devices worldwide \cite{williams2017pt1}. The UMOS technology is continuously supplanting and is often preferred over the traditional Double-diffused MOS (DMOS) technology due to advantages such as higher transistor cell density, uniform epitaxial current distribution, lower \textsc{on}-resistance, and reduced gate charge \cite{williams2017pt1, williams2017pt2}. \figureref{umos-dmos} illustrates the structural differences between DMOS and UMOS devices. However, the behavior of UMOS devices under ionizing radiation has not yet been extensively studied. Radiation including photons, neutrons, electrons protons, and heavy ions can damage these devices through total ionizing dose, displacement damages, or even charge deposition by single particle, the so-called Single-Event Effect (SEE) \cite{aguiar2024}. Destructive SEE failure modes in power MOSFETs include Single-Event Gate Rupture (SEGR) and Single-Event Burnout (SEB) \cite{sexton2003}. 

In terrestrial environments, SEB induced by atmospheric neutrons poses a critical challenge for modern power electronics reliability at ground level and flight altitudes \cite{normand1997, zhou2024}.
Although substantial research has been conducted on destructive radiation effects in DMOS transistors \cite{titus2013, liu2024}, relatively few studies have addressed power FETs with alternative technologies. 
Over the past decade, the lack of comparative studies on similarly rated UMOS and DMOS devices has been highlighted as a critical gap in the literature \cite{galloway2014}.
Existing research on SEB in silicon (Si)-based UMOSFETs has primarily focused on space applications, relying on computational simulations \cite{wang2013research, wang2013single, wang2014, wang2017} and limited experimental data \cite{wang2022, yu2022}.
An existing computational study concluded that Si UMOS devices might be more resistant to SEB than DMOSFETs \cite{wang2013research}. 
Detailed comparative SEB studies in silicon-carbide (SiC) UMOS and DMOS power FETs have been recently published, concluding higher SEB tolerance in UMOSFETs exposed to protons and neutrons \cite{martinella2021,martinella2023}. 
Despite these advances, detailed and comparative studies of SEB in Si-based UMOS and DMOS power FETs  are still lacking.

The apparent robustness of SiC UMOSFETs cannot be directly extrapolated to Si UMOSFETs, as the SEB mechanism in SiC devices is still under debate. It has been argued that the charge multiplication mechanisms responsible for SEB in Si MOSFETs may be suppressed in SiC devices \cite{martinella2021}.
Preliminary comparative studies on Si-based devices have revealed that UMOSFETs may prematurely exhibit enhanced charge multiplication effects, compared to DMOS counterparts, under monoenergetic fast neutron irradiation \cite{alberton2025dtneutrons, alberton2022radecs, alberton2024radecs}. 
These findings suggest that UMOSFETs might be more prone to SEB mode than previously predicted by computational simulations, emphasizing the need for further experimental investigations. 
Assessing potential failure risks in the UMOS technology under atmospheric neutrons is crucial before its incorporation in avionics systems and ground level applications demanding high reliability.
Moreover, the prominence of UMOS devices in modern power electronics and the ubiquity of atmospheric neutrons underscore the significance of the present investigation.
 
In this study, the relative vulnerability of Si-based UMOSFETs to SEBs induced by quasi-atmospheric neutrons compared to DMOSFET counterparts is experimentally assessed.
Failure rates are estimated from experimental data and implications for ground level and avionics applications are discussed. 
Simulations provide insights into the neutron-induced SEB mechanism and potential reinforcement improvements.

\section{Materials and Methods}
\begin{figure}[t]   
    \centering
    \includegraphics[width=0.48\textwidth]{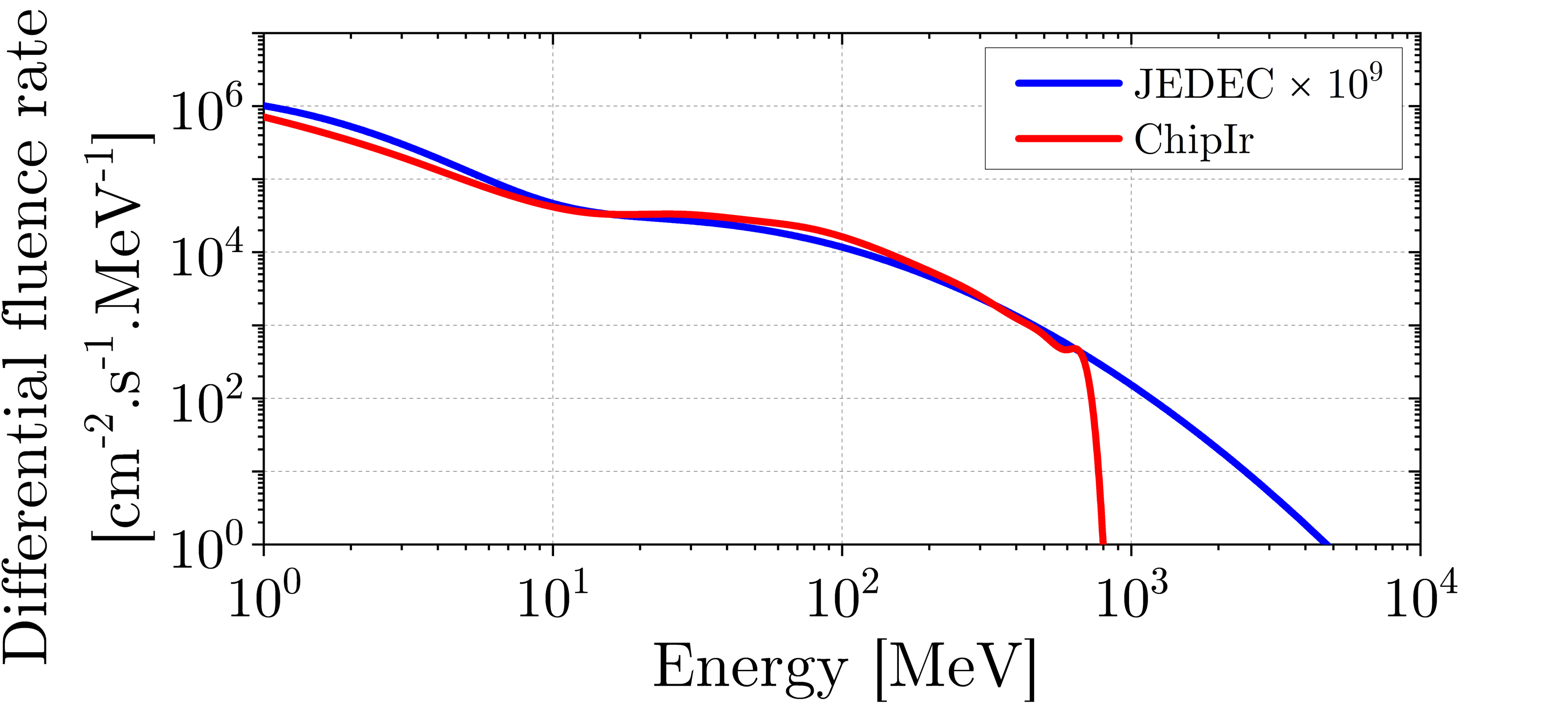}
    \caption[]{JEDEC and ChipIr neutron differential fluence rate as a function of neutron kinetic energy. After \cite{gordon2004} and \cite{chiesa2018}.}
    \label{fig:jedec-chipir}
\end{figure}

Currently, only a limited number of specialized spallation facilities worldwide are capable to provide neutron beams that replicate atmospheric neutron conditions for accelerated radiation testing.
Experimental investigations were conducted by using the ChipIr instrument, at the ISIS Neutron and Muon Source facility, UK. ChipIr provides an intense neutron beam with an energy spectrum similar to that measured in terrestrial environments under reference conditions \cite{cazzaniga2018, chiesa2018}, as specified by the JEDEC standard JESD89B (sea level, cutoff = 2.08 GV,  mid-level solar activity, outdoors) \cite{jedec-jesd89b, gordon2004}. \figureref{jedec-chipir} compares the neutron energy spectra of the quasi-atmospheric ChipIr beam and JEDEC atmospheric standard. 
Although the spectra are similar, the ChipIr spectrum is limited to energies below $800\; \mathrm{MeV}$ due to the maximum energy allowed by the ISIS synchrotron.

\begin{figure}[t]   
    \centering
    \includegraphics[width=0.43\textwidth]{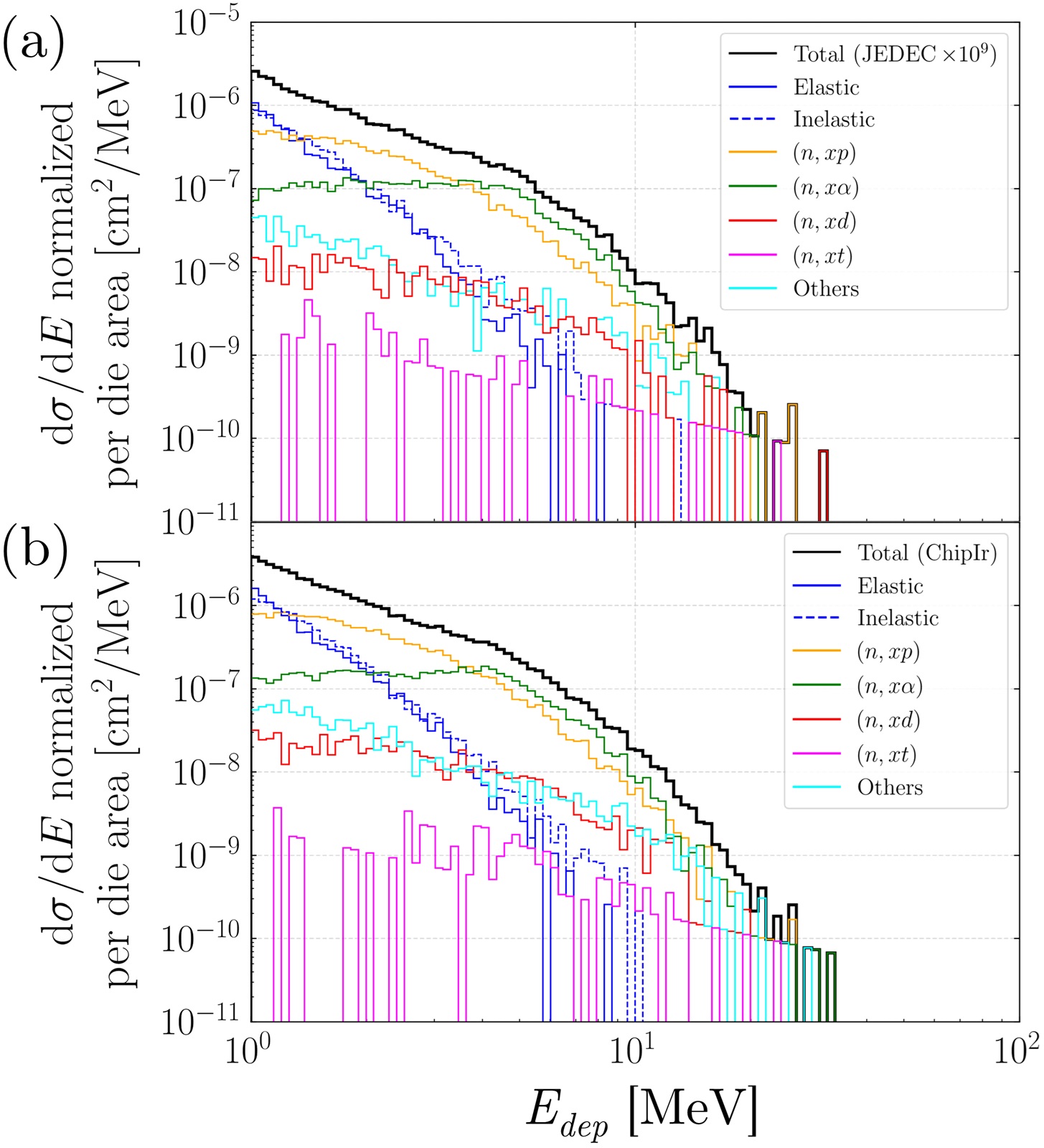}
    \caption[]{Differential cross section distributions of energy deposition within the epitaxial region of a generic $150\;\mathrm{V}$-rated MOSFET resulting from nuclear reactions induced by JEDEC and ChipIr neutrons. In our classification, ($n,xd$) and ($n,xt$) denote reactions producing at least one deuteron ($d$) and triton ($t$), respectively. Channels ($n,xp$) and ($n,x\alpha$) represent reactions in which protons ($p$) and alpha particles ($\alpha$), respectively, are the predominant ejectiles. Simulated using G4SEE \cite{lucsanyi2022}.}
    \label{fig:g4see}
\end{figure}

Computational simulations using the G4SEE toolkit \cite{lucsanyi2022} were conducted to verify the impact of the ISIS synchrotron energy cutoff on emulating JEDEC spectrum, and to identify the nuclear reaction secondaries most responsible for SEB occurrence. \figureref{g4see} presents simulations of the energy deposition distributions from neutron-induced secondary particles within the epitaxial region of a generic Si-based $150\;\mathrm{V}$ MOSFET. The hypothetical MOSFET was defined with a $5\;\textrm{\textmu m}$ Al metallization layer, a $13.5\;\textrm{\textmu m}$ epitaxial layer, and a $286.5\;\textrm{\textmu m}$ substrate. The similarity of deposited energy distributions per nuclear reaction channel confirms that the ISIS synchrotron energy cutoff has minimal impact on accurately emulating the atmospheric neutron environment with the ChipIr neutron beam. Additionally, technology computer-aided design (TCAD) simulations using ECORCE software \cite{michez2015} were conducted to analyze the experimental findings and support the interpretation of the underlying mechanisms of neutron-induced SEB in Si UMOSFETs. For simplicity, simulations were performed on generic $150\;\mathrm{V}$-rated UMOS and DMOS structures with representative parameters. Design parameters of the drift layer were estimated according to well-established semi-empirical expressions derived from Baliga's power law, whereas taking into account the influence of edge termination in practical power MOSFETs \cite{baliga2018}. 
Other design parameters were adapted from numerical examples presented in \cite{baliga2010} for similarly rated UMOS and DMOS devices.

\begin{figure}[t]   
    \centering
    \includegraphics[width=0.4\textwidth]{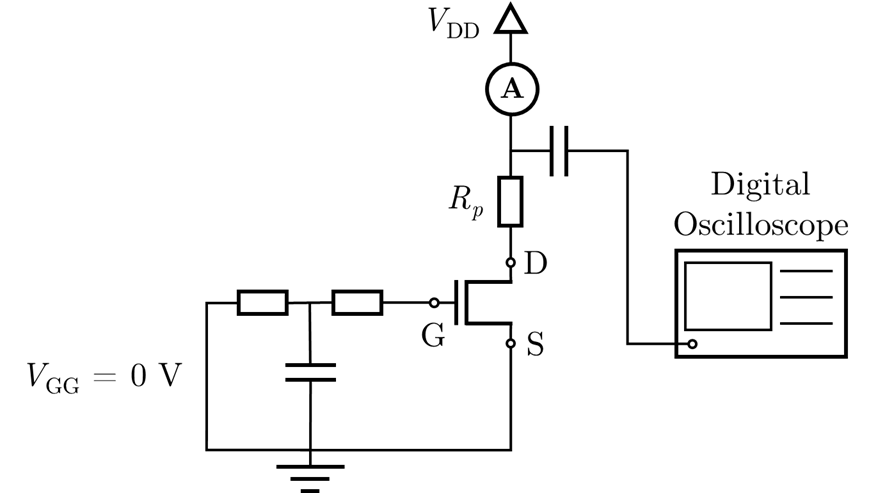}
    \caption[]{Schematic circuit diagram for protective SEB cross section measurements in power devices. Based on \cite{milstd750e}.}
    \label{fig:seb-circuit}
\end{figure}

Several n-type UMOS and DMOS power FETs with nominal voltage ratings of $40\,\mathrm{V}$, $60\,\mathrm{V}$, and $150\,\mathrm{V}$ were irradiated from the front side at neutron fluences of up to $10^{10} \, \mathrm{neutrons/cm}^2$.
The devices were irradiated as a function of the drain-source voltage ($V_{\mathrm{DS}}$), supplied by a source measure unit (Keithley 2410), while operating in the non-negative gate OFF-state regime ($V_\mathrm{GS}=0\;\mathrm{V}$ and $V_\mathrm{DS}>0\;\mathrm{V}$).
At least two devices of each part were tested following a standard method for protective evaluation of SEB by using the current limiting technique \cite{milstd750e}. 
The current-limiting technique is the primary method for protective SEB testing \cite{titus2013, milstd750e,liu2012}, enabling statistical assessment of reliability metrics, such as failure-in-time (FIT) rates and mean time to failure (MTTF) \cite{lee2011}, within a feasible experimental timeframe.
\figureref{seb-circuit} shows the schematic diagram of the SEB protective circuit, which basically consists of a $R$-$C$-$R$ network to prevent gate damage and a protection resistor on the drain to limit overcurrent \cite{titus2013}. SEB causes a transition from the OFF-state to a temporary ON-state, and the SEB-type pulses can be captured by using an oscilloscope. The adopted failure criterion is similar to that presented in \cite{alberton2022}, with the SEB signals monitored with a digital oscilloscope (InfiniiVision DSOx4024A) configured with a $-30\;\mathrm{mV}$ trigger level.

%

\section{Results and Discussion}
\begin{figure}[t]   
    \centering
    \includegraphics[width=0.4\textwidth]{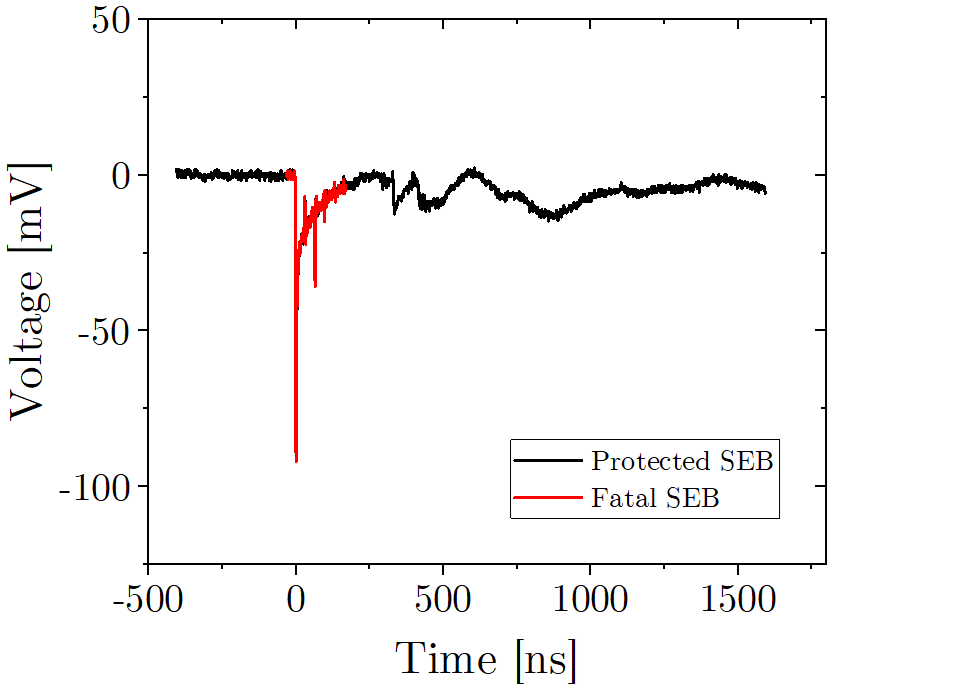}
    \caption[]{Representative SEB signals recorded on the oscilloscope. The black line represents a neutron-induced SEB successfully protected by the test circuit, allowing the device under test to remain operational and withstand high voltage. The red line represents a fatal SEB, where the protective circuitry fails to prevent drain-source damage, resulting in an $I_\mathrm{DSS}$ increase to the milliampere range.}
    \label{fig:seb-signals}
\end{figure}

Under the OFF-state bias conditions adopted in our study, SEB was verified to be the dominant neutron-induced destructive failure mode. Except for $150\;\mathrm{V}$ UMOSFETs, all other devices exhibited successful protection against SEB, mantaining low drain-source leakage current ($I_\mathrm{DSS}$) and nominal gate-source threshold voltages ($V_{th}$).
However, for the $150\;\mathrm{V}$ UMOSFETs operating at high voltage, the protective circuitry was occasionally ineffective, resulting in sudden increases in $I_\mathrm{DSS}$ to the milliampere range (fatal events). Post-irradiation characterization confirmed permanent drain-source damage in some $150\;\mathrm{V}$ UMOSFETs, with the gate remaining functional, although $V_{th}$ was increased. In other cases, both drain and gate damages were observed, suggesting combined SEB and SEGR occurrence. \figureref{seb-signals} compares recorded signals of a successfully protected SEB (black) with a fatal event (red), highlighting that both signals are nearly identical except for the drain-source damage in the latter. In cases where the circuit failed to protect neutron-induced destructive failures, the damaged devices were replaced and voltage levels reset before resuming measurements.

\begin{figure}[t]   
    \centering
    \includegraphics[width=0.46\textwidth]{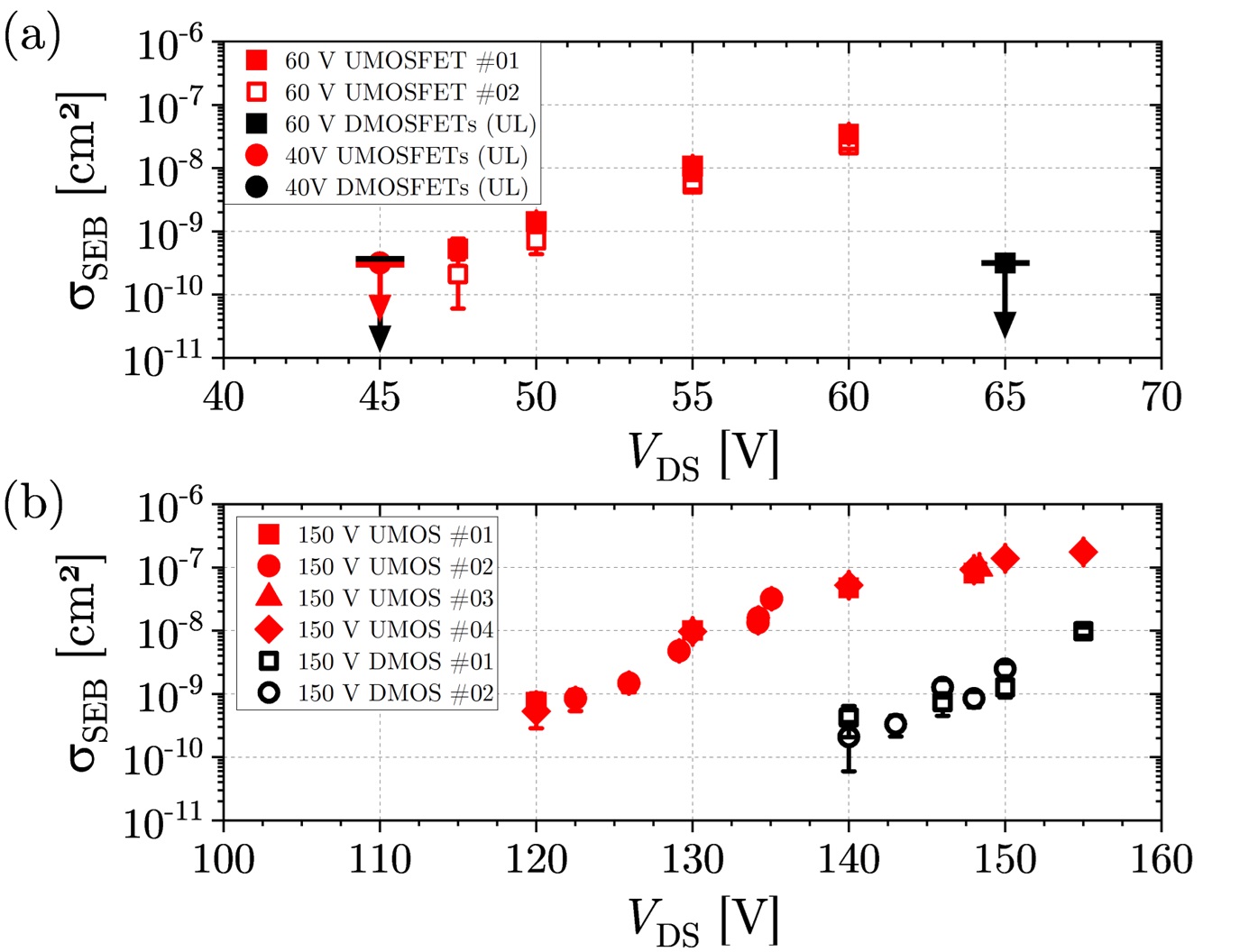}
    \caption[]{SEB cross section measurements of $40\;\mathrm{V}$-, $60\;\mathrm{V}$-, and $150\;\mathrm{V}$-rated UMOS and DMOS power FETs irradiated with quasi-atmospheric neutrons at ChipIr, UK. The arrows denote the upper limits (UL) at the 90\% confidence level where no SEBs were observed. (a) 40 V and 60 V-rated devices. (b) $150\;\mathrm{V}$-rated devices. Tested devices are identified by enumeration label.}
    \label{fig:xsec}
\end{figure}

The SEB sensitivity of the devices under test (DUTs) was evaluated in terms of the SEB cross section \cite{liu2024}:
\begin{equation}
\sigma_\mathrm{SEB} = \frac{N_\mathrm{SEB}}{\Phi} \, ,
\end{equation}
defined as the ratio of the total number of detected SEB signals ($N_\mathrm{SEB}$) to the particle beam fluence ($\Phi$) for each test run. 
The SEB cross section quantifies the likelihood of SEB occurrence, reflecting the cross-sectional area of the device's sensitive volume that effectively contributes to triggering SEBs.
As shown in \figref{xsec}, the $\sigma_\mathrm{SEB}$ measurements as a function of $V_\mathrm{DS}$ are plotted for SEBs induced by quasi-atmospheric neutrons in UMOS and DMOS power FETs. No SEBs were observed in either $40\,\mathrm{V}$ UMOS, $40\, \mathrm{V}$ DMOS, or $60\,\mathrm{V}$ DMOS devices, even when irradiated at drain-source voltages very close to their actual breakdown voltages ($BV_\mathrm{DS}$). Consequently, 90\% confidence level upper limits (UL) for $\sigma_\mathrm{SEB}$ are indicated by arrows in \figref[(a)]{xsec} for these cases. In contrast, SEBs were observed in $60 \, \mathrm{V}$ UMOSFETs at voltages as low as $V_\mathrm{DS}=47.5 \, \mathrm{V}$. The maximum SEB cross section measured for the $60 \, \mathrm{V}$ UMOSFET was approximately two orders of magnitude higher than the upper limit estimated for the $60 \, \mathrm{V}$ DMOSFET. \figureref[(b)]{xsec} shows a similar trend for $150 \, \mathrm{V}$ devices, with SEBs being triggered from approximately $V_\mathrm{DS}= 120 \, \mathrm{V}$ in UMOSFETs, compared to about $V_\mathrm{DS}= 140 \, \mathrm{V}$ in DMOSFETs. Comparatively, SEB cross sections are higher for $150 \, \mathrm{V}$ UMOSFETs.

\begin{figure}[t]   
    \centering
    \includegraphics[width=0.45\textwidth]{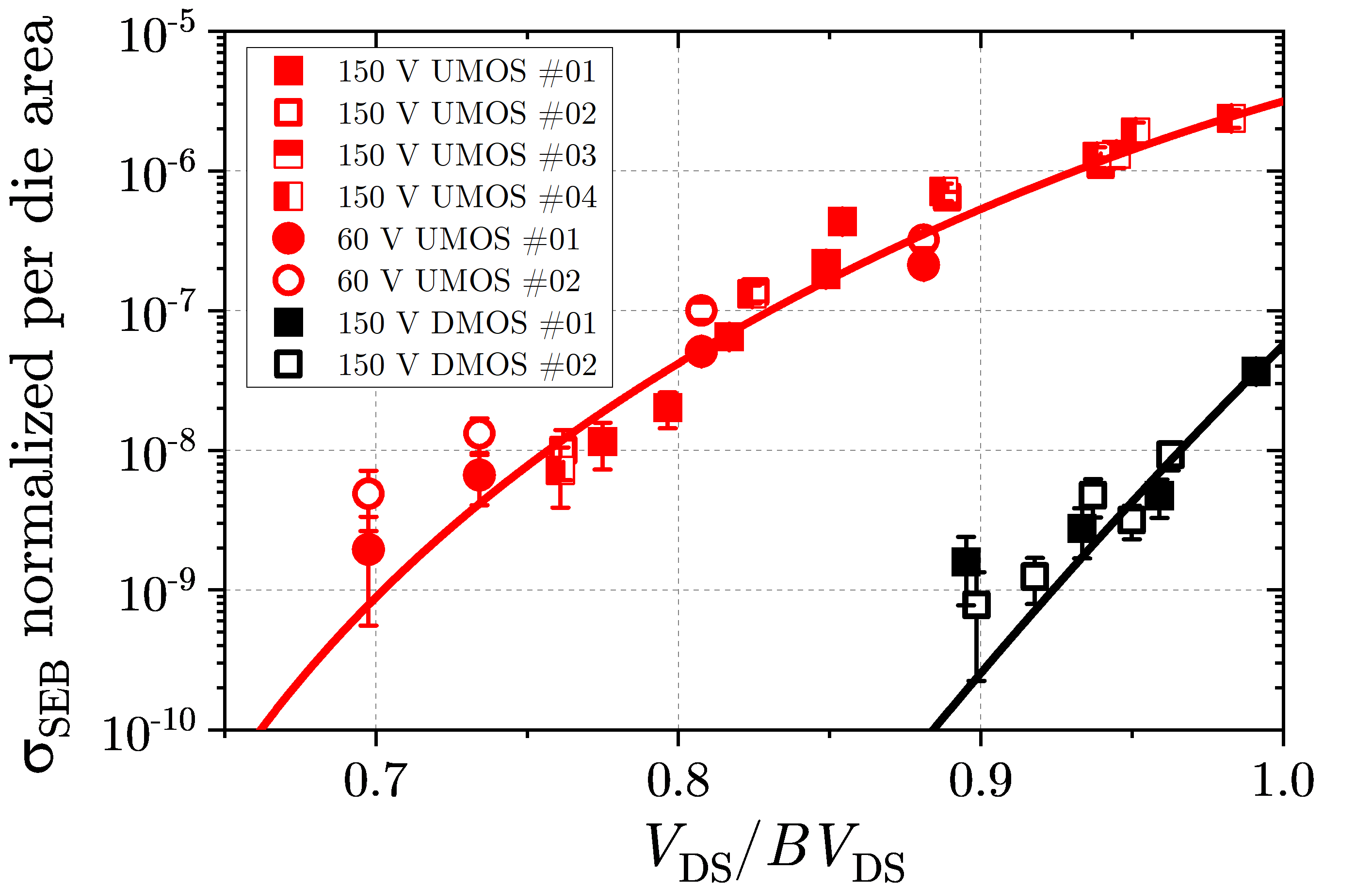}
    \caption[]{SEB cross section measurements normalized per die area as a function of the applied drain-source voltage normalized per actual breakdown voltage, according to precision measurements. Tested devices are identified by enumeration label. Solid curves represent the Weibull curve fits to experimental data.}
    \label{fig:norm-xsec}
\end{figure}

In order to accurately determine safe operating voltages across technologies and directly compare their SEB susceptibilities, it is advantageous to evaluate the $\sigma_\mathrm{SEB}$ values normalized per die area as a function of the $V_\mathrm{DS}$ values normalized per the actual breakdown voltage $BV_\mathrm{DS}$. \figureref{norm-xsec} presents the dependence of the normalized SEB cross sections on the applied voltage ratio of the DUTs. By fitting Weibull statistical distributions on experimental data, the SEB threshold voltages are determined. Results reveal that SEB can initiate in UMOSFETs at voltage ratios as low as approximately 66\% of $BV_\mathrm{DS}$, compared to a threshold of around 88\% for DMOSFETs. As shown in \figref{norm-xsec}, the typical security derating level of 75\% \cite{sahu2003} is insufficient to completely prevent SEB in UMOSFET parts rated as low as 60 V. Furthermore, near the maximum operational voltage level, normalized SEB cross sections of UMOS and DMOS differ by two orders of magnitude, confirming the intrinsic vulnerability of UMOSFETs regardless of die area dimensions.

Assuming the ChipIr neutron spectrum closely resembles the JEDEC reference spectrum, SEB failure rates induced by atmospheric neutrons can be estimated for any location and condition on Earth by re-scaling the reference neutron fluence rate \cite{gordon2004, jedec-jesd89b}. 
Under reference conditions, the average Single-Event Rate (SER) can be computed as follows \cite{jedec-jesd89b}:
\begin{equation}
\mathrm{SER} = \int \sigma(E) \, \frac{d\phi(E)}{dE} \, dE = \sigma_\mathrm{SEB} \, \phi_{ref} \, ,
\end{equation}
where $d\phi/dE$ is the reference neutron differential fluence rate, as shown in \figref{jedec-chipir}, and $\phi_{ref}=12.95 \;\mathrm{neutrons.cm}^{-2}.\mathrm{h}^{-1}$ is the reference neutron fluence rate, according to the JEDEC standard \cite{jedec-jesd89b}.
For instance, when expressed in conventional FIT units (failures per $10^9\,\mathrm{h}$), the estimated SEB failure rates for the $60\; \mathrm{V}$ UMOS, $150\; \mathrm{V}$ DMOS, and $150\; \mathrm{V}$ UMOS devices operating at nominal voltage and reference sea level are $450(60)\; \mathrm{FIT}$, $32(7)\; \mathrm{FIT}$, and $1.8(3)\times 10^3\; \mathrm{FIT}$, respectively. The UMOS devices exceed the specific $100\; \mathrm{FIT}$ reliability requirement stipulated by the JEDEC standard JEP151A \cite{jedec-jep151a}, whereas the $150\; \mathrm{V}$ DMOS does not. 
The MTTF for SEB occurrence in these $60\;\mathrm{V}$ and $150\;\mathrm{V}$ UMOSFETs operating at nominal voltage and sea level are $250(30)\; \mathrm{years}$ and $63(11)\; \mathrm{years}$, respectively.
At a commercial aviation altitude of $12\; \mathrm{km}$, where neutron fluence rate increases by about 500 times \cite{gordon2004}, the corresponding MTTF values reduce to $0.51(7)\; \mathrm{year}$ and $0.127(22)\; \mathrm{year}$, respectively. According to JEDEC standards, these UMOS failure rates are unacceptable for avionics and inadequate for traction and automotive applications at ground level.

The SEB mechanism in Si-based power devices involves impact ionization and carrier tunneling, that are strongly dependent on the local electric field intensity \cite{kuboyama2004}. This fact suggest that the UMOS vulnerability is likely due to electric field stresses present near its trench corner (see \figref{umos-dmos}), which are substantially more intense than those found in similarly rated DMOS structures \cite{baliga2010}.
These intense fields can enhance multiplication of charge carriers produced by neutron-induced secondary particles, triggering avalanche multiplication, activating the parasitic bipolar junction transistor (BTJ) and, under critical conditions, ultimately lead to SEB. The neutron-induced SEB can be interpreted in terms of the Egawa effect \cite{egawa1966} and current-induced avalanche breakdown \cite{wrobel1985}. According to the Poisson equation, the electric field ($\mathcal{E}$) gradient is affected by the effective charge carrier distribution within the epitaxial region \cite{wrobel1985}:
\begin{equation}
\label{eq:poisson}
\frac{d\mathcal{E}}{dx} = \frac{q}{\varepsilon} [N_d - n(x)] = \frac{q}{\varepsilon} \left( N_{epi} - \frac{J}{q\, v_{sat}} \right) \, ,
\end{equation}
where $N_d=N_{epi}$ is the epitaxial region doping concentration, and $J\cong q\, v_{sat}\, n(x)$ is the drift current density of carriers induced by secondaries. 
Under avalanche, $n(x)$ increases due to ionization impact. When $n(x) \gg N_{epi}$, the electric field peak is shifted toward the epitaxial-substrate interface due to Egawa effect \cite{tambone2024}. Whether the maximum electric field intensity exceeds the semiconductor critical field ($\mathcal{E}_{max} > \mathcal{E}_{crit}$), a SEB occurs \cite{wrobel1985}. For silicon, $\mathcal{E}_{crit}\approx 3\times 10^5 \;\mathrm{V/cm}$ \cite{wrobel1985}.

\begin{figure}[t]   
    \centering
    \includegraphics[width=0.45\textwidth]{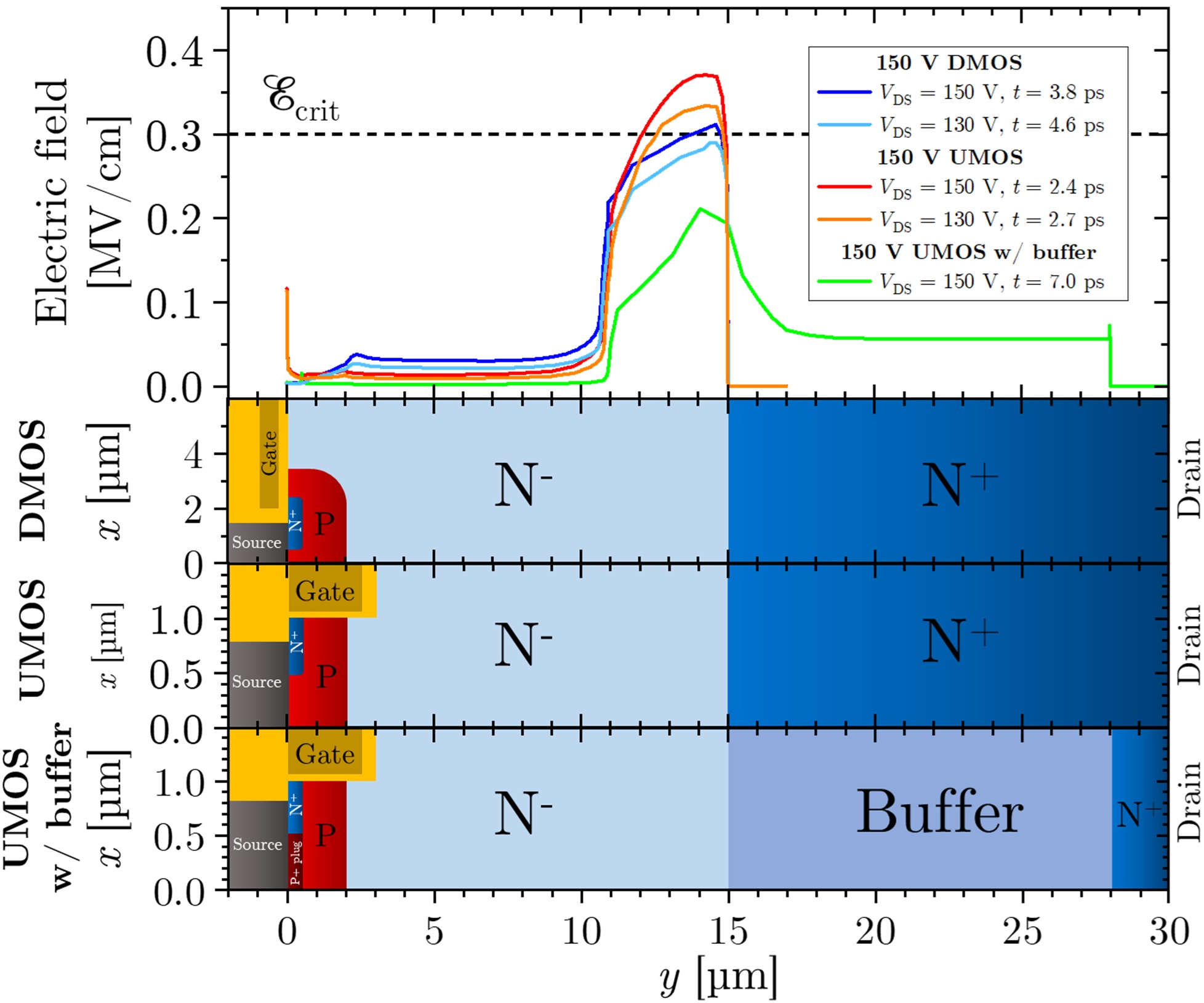}
    \caption[]{Spatial distributions of the peak electric field after neutron-induced SEE in $150\;\mathrm{V}$-rated DMOS, UMOS, and UMOS with buffer layer devices. Based on G4SEE simulation, a secondary $25.7\;\mathrm{MeV}$ $^{27}\mathrm{Al}$ ion, created in $x=0.5\;\textrm{\textmu m}$, $y=0\;\textrm{\textmu m}$, is emitted in the $y$ direction at $t=0.2\;\mathrm{ps}$. Given the minimal energy deposition of the accompanying $128\;\mathrm{MeV}$ proton, only the $^{27}\mathrm{Al}$ ion is considered in simulations using ECORCE \cite{michez2015}.}
    \label{fig:ecorce}
\end{figure}

Coupled G4SEE and TCAD simulations were conducted to exemplify the previously mentioned mechanism, elucidating the experimentally observed UMOSFET vulnerability.
Based on \figref{g4see}, post-processing analysis indicates that high-energy protons generally escape the epitaxial volume with minimal energy deposition, whereas the most energetic events usually involve residual heavy nuclei and alpha particles. For instance, one of the highest-energy events was initiated by a $243.2\;\mathrm{MeV}$ primary neutron, which produced a $70.1\;\mathrm{MeV}$ secondary neutron, a $128\;\mathrm{MeV}$ proton, and a $25.7\; \mathrm{MeV}$ $^{27}\mathrm{Al}$ ion. These secondary particles deposited energies of $0\;\mathrm{keV}$, $5\;\mathrm{keV}$, and $23.9\;\mathrm{MeV}$, respectively. 
Due to the negligible energy deposition by high-energy protons, this particular event has been selected as the simplest representative worst-case scenario for atmospheric neutron-induced SEB, where only the $^{27}\mathrm{Al}$ ion can be considered in the TCAD simulations for convenience.
The simulated doping profiles followed the geometry shown in \figref{ecorce}, with doping concentrations defined as specified in \tabref{ecorce}.
\figureref{ecorce} shows the distributions of peak electric field after neutron-induced SEE in $150\;\mathrm{V}$-rated devices for the representative case described earlier. In both UMOS and DMOS operating at $V_\mathrm{DS}=150\;\mathrm{V}$, the electric field peak shifts toward epitaxial-substrate interface, exceeding $\mathcal{E}_{crit}$ and initiating SEB. 
Complementary TCAD simulations at $V_\mathrm{DS}=130\;\mathrm{V}$ align with the experimental results shown in \figref[(b)]{xsec}, where neutron-induced SEB occurs in the $150\;\mathrm{V}$ UMOS but not in the $150\;\mathrm{V}$ DMOS under the same condition.
Being consistent with experimental results and the analytical interpretation (\ref{eq:poisson}), the TCAD simulations corroborate that UMOS devices are indeed more vulnerable to SEBs.

\begin{table}[t!]
   	\centering 
	\begin{threeparttable}
   	\caption[]{Doping concentrations, in units of $\mathrm{cm}^{-3}$, adopted in TCAD simulations with ECORCE software \cite{michez2015}.}
   	\label{tab:ecorce}
	\small 
	\begin{tabular}{@{}*4c@{}} 
   	\toprule[\heavyrulewidth]\toprule[\heavyrulewidth]
	\textbf{Region}  & 
	\textbf{150 V DMOS}  &
	\textbf{150 V UMOS} &
	\makecell{ \textbf{150 V UMOS} \\ \textbf{with buffer} }\\ 
   	\midrule     	
	\multirow{1}{5em}{\centering $n^+$-source} & $1.00\times 10^{19}$ & $1.00\times 10^{19}$ & $1.00\times 10^{19}$ \\
	\addlinespace[0.1cm]	
	\hline
	\addlinespace[0.1cm]		
	\multirow{1}{5em}{\centering $n^+$-drain} & $1.00\times 10^{20}$ & $1.00\times 10^{20}$ & $1.00\times 10^{20}$ \\
	\addlinespace[0.1cm]	
	\hline
	\addlinespace[0.1cm]
	\multirow{1}{5em}{\centering $n^-$-drift} & $1.47\times 10^{15}$ & $1.47\times 10^{15}$ & $1.47\times 10^{15}$ \\
	\addlinespace[0.1cm]	
	\hline
	\addlinespace[0.1cm]
		\multirow{1}{5em}{\centering $p$-base} & $1.00\times 10^{17}$ & $1.00\times 10^{17}$ & $1.00\times 10^{17}$ \\
	\addlinespace[0.1cm]	
	\hline
	\addlinespace[0.1cm]	
	\multirow{1}{5em}{\centering $p^+$-plug} & - & - & $1.00\times 10^{19}$ \\
	\addlinespace[0.1cm]
	\hline
	\addlinespace[0.1cm]	
	\multirow{1}{5em}{\centering $n$-buffer } & - & - & $1.00\times 10^{16}$ \\

   	\bottomrule[\heavyrulewidth] 
   	\end{tabular}  
\end{threeparttable} 
\newline
\end{table}

Experimental results demonstrate that current UMOSFET designs are inadequate for high-reliability applications in atmospheric radiation environments.
However, simulations suggest that incorporating a properly doped buffer layer \cite{liu2007} can mitigate neutron-induced SEB by reducing peak electric fields after a particle strike, as shown in \figref{ecorce}. It is shown that the buffer layer reduces the electric field peak to safe levels after particle strike, effectively preventing premature SEB occurrence in the UMOSFET structure.
Previous studies have shown that buffer layers significantly improve SEB performance of DMOSFETs \cite{liu2006, liu2007, ferlet2010}. Recent investigations also confirm that buffer layers enhance SEB hardness in split-gate UMOSFETs under 2 GeV tantalum irradiation \cite{yu2022}. Although such high-energy heavy-ion irradiation conditions are not representative of nuclear reaction secondaries in atmospheric environments, these results suggest that similar radiation-hardness can be achieved in next-generation UMOSFETs incorporating buffer layers. Further research is required to optimize buffer layer design for power UMOSFETs and evaluate its effectiveness against neutron-induced SEB in atmospheric environments.

\begin{figure}[t]   
    \centering
    \includegraphics[width=0.4\textwidth]{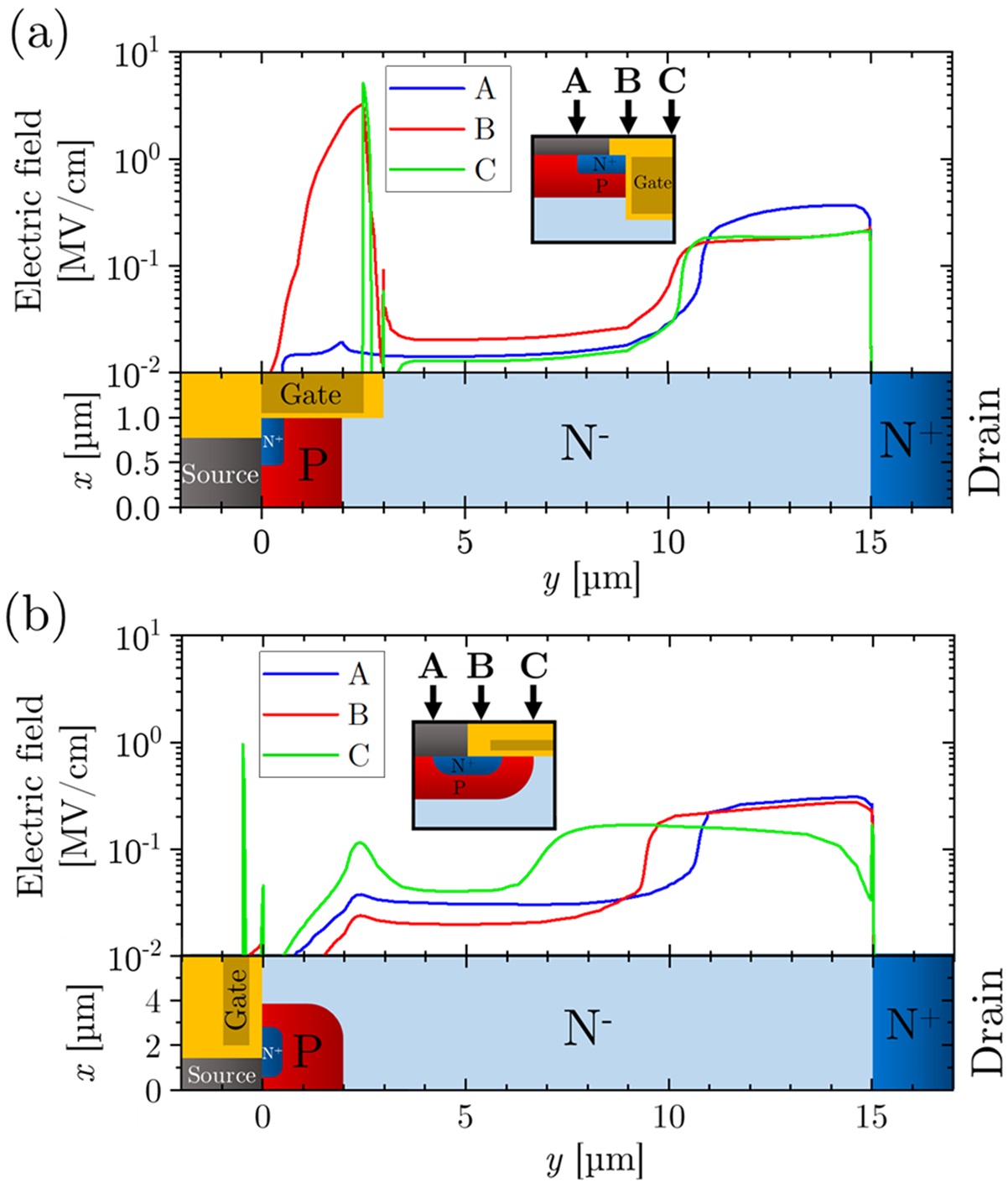}
    \caption[]{Spatial distributions of the peak electric field in $150\;\mathrm{V}$-rated (a) UMOS and (b) DMOS devices following secondary ion impacts at distinct positions. Graphs presented in logarithmic scale. Simulated using ECORCE \cite{michez2015}.}
    \label{fig:ecorce-oxide}
\end{figure}

\begin{figure}[t]   
    \centering
    \includegraphics[width=0.45\textwidth]{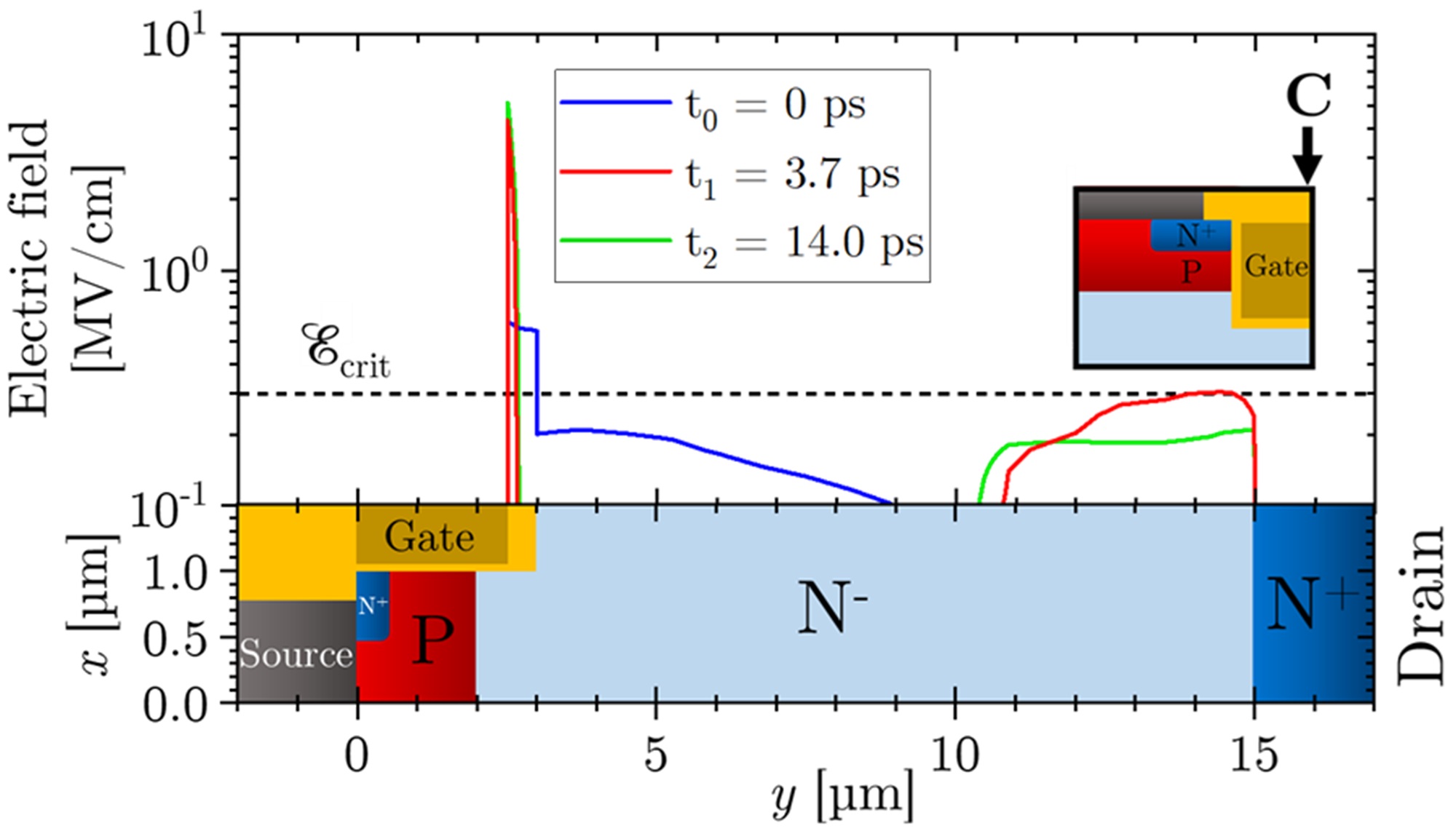}
    \caption[]{Temporal evolution of the electric field distributions in $150\;\mathrm{V}$-rated UMOS following secondary ion impact over the trench gate column (position C) at $t=0.2\;\mathrm{ps}$. Simulated using ECORCE \cite{michez2015}.}
    \label{fig:ecorce-temporal}
\end{figure}

\begin{figure}[t!]   
    \centering
    \includegraphics[width=0.45\textwidth]{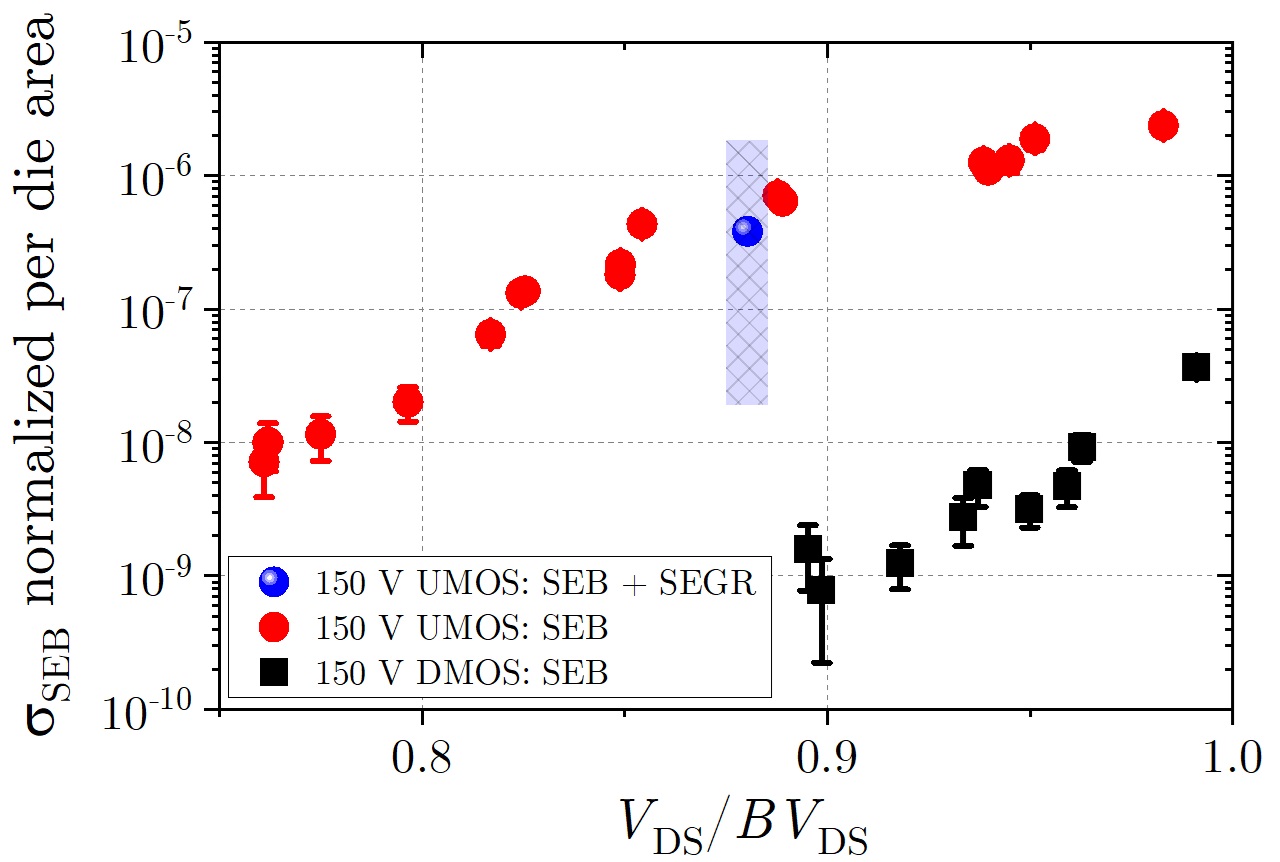}
    \caption[]{Normalized cross section of combined SEB and SEGR failure in $150\;\mathrm{V}$-rated UMOSFETs compared to the pure SEB cross section measurements. The blue region indicates the 90\% confidence interval.}
    \label{fig:norm-seb-segr}
\end{figure}

Although not the primary focus of this work, other failure mechanisms, such as gate degradation \cite{tambone2024, grasser2014} or SEGR \cite{sexton2003}, can affect the reliability of MOSFETs. Nevertheless, some insights can be drawn relating these failure modes to our findings on the vulnerability of UMOSFETs to atmospheric neutrons. Recent experimental results have demonstrated that monoenergetic fast neutrons from Deuterium-Tritium (D-T) neutron generators can prematurely induce avalanche multiplication in Si-based power UMOSFETs \cite{alberton2025dtneutrons, alberton2022radecs, alberton2024radecs}. In atmospheric environments, due to the greater ionization coefficient of electrons compared to holes, it is likely that ionized electrons produced by nuclear reaction secondaries can enhance the risk of gate degradation through hot-carrier injection during avalanche conditions \cite{williams2017pt2}. Furthermore, the increased risk of SEGR is illustrated in \figref{ecorce-oxide}, which presents TCAD simulations of peak electric field distributions in generic $150\;\mathrm{V}$-rated UMOS and DMOS devices for different secondary ion impact positions. Depending on the location of the nuclear reaction secondary incidence, intense electric field spikes can form across the gate oxide. Whether the peak electric field exceeds the dielectric strength of SiO$_2$ (approximately 10 MV/cm \cite{sze2021}), SEGR can occur, leading to catastrophic gate oxide rupture. For instance, \figref{ecorce-oxide} indicates that electric field intensities exceeding 5 MV/cm can occur across the gate oxide of the $150\;\mathrm{V}$ UMOSFET, which is of the same order of magnitude as the dielectric strength of SiO$_2$. In contrast, peak electric fields lower than 1 MV/cm are observed across the gate oxide of the $150\;\mathrm{V}$ DMOSFET, highlighting the increased vulnerability of UMOSFETs to the neutron-induced SEGR failure compared to DMOSFETs. Similar to how intense electric fields in UMOSFETs promote SEB, it is plausible that neutron-induced SEEs followed by avalanche multiplication can also, in extreme cases, promote SEGR. \figureref{ecorce-temporal} shows the temporal evolution of the electric field distribution in the $150\;\mathrm{V}$ UMOSFET for secondary ion impact occurring over the trench gate column at $t=0.2\;\mathrm{ps}$. Simulations reveal that the electric field in the gate oxide continues to increase after the peak electric field in the drift-substrate junction exceeds the critical field of Si at approximately $3.7\;\mathrm{ps}$. These results suggest that the dynamics of combined SEB and SEGR failures, as experimentally observed in some $150\;\mathrm{V}$ UMOSFETs, are primarily governed by the SEB mechanism, which is triggered first. This conclusion aligns with experimental data presented in \figref{norm-seb-segr}, showing the average cross section of combined SEB and SEGR failure for $150\;\mathrm{V}$ UMOSFETs. The blue region indicates the 90\% confidence interval. Although representing distinct failure mechanisms, the normalized cross section of combined SEB and SEGR failure is statistically compatible with pure SEB cross sections for $150\;\mathrm{V}$ UMOSFETs.

\section{Conclusion}
Unlike SiC-based power FETs and contrary to early computational predictions, experimental results demonstrate that Si-based UMOSFETs are generally more vulnerable and prematurely experience SEBs in terrestrial radiation environments compared to DMOSFET counterparts. 
This increased vulnerability is likely attributed to the high local electric fields near the trench corners of UMOSFETs, which favor avalanche multiplication that ultimately initiate SEBs.
Failure rates at both reference sea level and flight altitudes indicate that current UMOSFETs designs are unsuitable for high-reliability atmospheric applications.
Nevertheless, simulations suggest that incorporating a buffer layer in next-generation UMOSFETs can mitigate SEBs induced by atmospheric neutrons, improving device reliability.
Experimental and computational findings indicate that SEB is the primary destructive failure mechanism in Si UMOSFETs operating in non-negative OFF-state regime, whereas SEGR may occur as a secondary effect in combination with SEB under extreme conditions.

\label{Bibliography}
\bibliographystyle{ieeetr}
\bibliography{Bibliography}

\begin{thebibliography}{10}
\providecommand{\url}[1]{#1}
\csname url@samestyle\endcsname
\providecommand{\newblock}{\relax}
\providecommand{\bibinfo}[2]{#2}
\providecommand{\BIBentrySTDinterwordspacing}{\spaceskip=0pt\relax}
\providecommand{\BIBentryALTinterwordstretchfactor}{4}
\providecommand{\BIBentryALTinterwordspacing}{\spaceskip=\fontdimen2\font plus
\BIBentryALTinterwordstretchfactor\fontdimen3\font minus
  \fontdimen4\font\relax}
\providecommand{\BIBforeignlanguage}[2]{{%
\expandafter\ifx\csname l@#1\endcsname\relax
\typeout{** WARNING: IEEEtran.bst: No hyphenation pattern has been}%
\typeout{** loaded for the language `#1'. Using the pattern for}%
\typeout{** the default language instead.}%
\else
\language=\csname l@#1\endcsname
\fi
#2}}
\providecommand{\BIBdecl}{\relax}
\BIBdecl

\bibitem{williams2017pt1}
R.~K. Williams, M.~N. Darwish, R.~A. Blanchard, R.~Siemieniec, P.~Rutter, and
  Y.~Kawaguchi, ``The trench power {MOSFET}: {P}art {I}—{H}istory,
  technology, and prospects,'' \emph{IEEE Trans. Electron Devices}, vol.~64,
  no.~3, pp. 674--691, Mar. 2017, doi:
  \href{https://doi.org/10.1109/TED.2017.2653239}{10.1109/TED.2017.2653239}.

\bibitem{williams2017pt2}
------, ``The trench power {MOSFET}—{P}art {II}: {A}pplication specific
  {VDMOS}, {LDMOS}, packaging, and reliability,'' \emph{IEEE Trans. Electron
  Devices}, vol.~64, no.~3, pp. 692--712, Mar. 2017, doi:
  \href{https://doi.org/10.1109/TED.2017.2655149}{10.1109/TED.2017.2655149}.

\bibitem{aguiar2024}
V.~A.~P. Aguiar, N.~H. Medina, N.~Added, S.~G. Alberton, E.~L.~A. Macchione,
  M.~A. Guazzelli, M.~A.~A. Melo, J.~A. Oliveira, R.~C. Giacomini, F.~R.
  Aguirre, P.~R.~P. Allegro, H.~C.~S. Zaggato, and I.~J. Sayeg, ``Evaluation of
  funnel models on calculation of ion-induced collected charge,'' \emph{IEEE
  Trans. Electron Devices}, vol.~72, no.~1, pp. 31--36, Jan. 2025, doi:
  \href{https://doi.org/10.1109/TED.2024.3497927}{10.1109/TED.2024.3497927}.

\bibitem{sexton2003}
F.~W. Sexton, ``Destructive single-event effects in semiconductor devices and
  {IC}s,'' \emph{IEEE Trans. Nucl. Sci.}, vol.~50, no.~3, pp. 603--621, Jun.
  2003, doi:
  \href{https://doi.org/10.1109/TNS.2003.813137}{10.1109/TNS.2003.813137}.

\bibitem{normand1997}
E.~Normand, J.~L. Wert, D.~L. Oberg, P.~Majewski, P.~Voss, and S.~Wender,
  ``Neutron-induced single event burnout in high voltage electronics,''
  \emph{IEEE Trans. Nucl. Sci.}, vol.~44, no.~6, pp. 2358--2366, Dec. 1997,
  doi: \href{https://doi.org/10.1109/23.659062}{10.1109/23.659062}.

\bibitem{zhou2024}
F.~Zhou, L.~Mo, Z.~Hu, Q.~Yu, C.~Zou, W.~Xu, F.~Ren, D.~Zhou, D.~Chen,
  Y.~Zheng, R.~Zhang, and H.~Lu, ``1.5-k{V} {A}l{G}a{N}/{G}a{N} {MIS}-{HEMT}
  with 3-{D} stacking pad-connected schottky structure demonstrating radiation
  robustness against atmospheric neutrons,'' \emph{IEEE Electron Device Lett.},
  vol.~45, no.~7, pp. 1129--1132, Jul. 2024, doi:
  \href{https://doi.org/10.1109/LED.2024.3403682}{10.1109/LED.2024.3403682}.

\bibitem{titus2013}
J.~L. Titus, ``An updated perspective of single event gate rupture and single
  event burnout in power {MOSFET}s,'' \emph{IEEE Trans. Nucl. Sci.}, vol.~60,
  no.~3, pp. 1912--1928, Jun. 2013, doi:
  \href{https://doi.org/10.1109/TNS.2013.2252194}{10.1109/TNS.2013.2252194}.

\bibitem{liu2024}
F.-K. Liu, Z.-L. Liu, and X.-J. Li, ``Impact of $^{60}${C}o-$\gamma$
  irradiation pre-treatment on single-event burnout in {N}-channel power
  {VDMOS} transistors,'' \emph{IEEE Electron Device Lett.}, vol.~45, no.~7, pp.
  1105--1108, Jul. 2024, doi:
  \href{https://doi.org/10.1109/LED.2024.3403570}{10.1109/LED.2024.3403570}.

\bibitem{galloway2014}
K.~F. Galloway, ``A brief review of heavy-ion radiation degradation and failure
  of silicon {UMOS} power transistors,'' \emph{Electronics}, vol.~3, no.~4, pp.
  582--593, Sep. 2014, doi:
  \href{https://doi.org/10.3390/electronics3040582}{10.3390/electronics3040582}.

\bibitem{wang2013research}
Y.~Wang, Y.~Zhang, and C.~Yu, ``Research of single-event burnout in power
  {UMOSFET}s,'' \emph{IEEE Trans. Electron Devices}, vol.~60, no.~2, pp.
  887--892, Feb. 2013, doi:
  \href{https://doi.org/10.1109/TED.2012.2234126}{10.1109/TED.2012.2234126}.

\bibitem{wang2013single}
Y.~Wang, Y.~Zhang, L.-G. Wang, and C.~Yu, ``Single-event burnout hardening of
  power {UMOSFET}s with optimized structure,'' \emph{IEEE Trans. Electron
  Devices}, vol.~60, no.~6, pp. 2001--2007, Jun. 2013, doi:
  \href{https://doi.org/10.1109/TED.2013.2256426}{10.1109/TED.2013.2256426}.

\bibitem{wang2014}
Y.~Wang, C.-H. Yu, Z.~Dou, and W.~Xue, ``Single-event burnout hardening of
  power {UMOSFET}s with integrated schottky diode,'' \emph{IEEE Trans. Electron
  Devices}, vol.~61, no.~5, pp. 1464--1469, 2014, doi:
  \href{https://doi.org/10.1109/TED.2014.2312948}{10.1109/TED.2014.2312948}.

\bibitem{wang2017}
Y.~Wang, C.-H. Yu, M.-S. Li, F.~Cao, and Y.-J. Liu, ``High-performance
  split-gate-enhanced {UMOSFET} with dual channels,'' \emph{IEEE Trans.
  Electron Devices}, vol.~64, no.~4, pp. 1455--1460, Apr. 2017, doi:
  \href{https://doi.org/10.1109/TED.2017.2665589}{10.1109/TED.2017.2665589}.

\bibitem{wang2022}
Y.~Wang, C.-H. Yu, X.-J. Li, and J.-Q. Yang, ``A comparative study on heavy-ion
  irradiation impact on p-channel and n-channel power {UMOSFET}s,'' \emph{IEEE
  Trans. Nucl. Sci.}, vol.~69, no.~6, pp. 1249--1256, Jun. 2022, doi:
  \href{https://doi.org/10.1109/TNS.2022.3175954}{10.1109/TNS.2022.3175954}.

\bibitem{yu2022}
C.-H. Yu, Y.~Wang, M.-T. Bao, X.-J. Li, J.-Q. Yang, and F.~Cao, ``Impact of
  heavy-ion irradiation in an 80-{V} radiation-hardened split-gate trench power
  {UMOSFET},'' \emph{IEEE Trans. Electron Devices}, vol.~69, no.~2, pp.
  664--668, Feb. 2022, doi:
  \href{https://doi.org/10.1109/TED.2021.3135369}{10.1109/TED.2021.3135369}.

\bibitem{martinella2021}
C.~Martinella, R.~G. Alia, R.~Stark, A.~Coronetti, C.~Cazzaniga, M.~Kastriotou,
  Y.~Kadi, R.~Gaillard, U.~Grossner, and A.~Javanainen, ``Impact of terrestrial
  neutrons on the reliability of {S}i{C} {VD}-{MOSFET} technologies,''
  \emph{IEEE Trans. Nucl. Sci.}, vol.~68, no.~5, pp. 634--641, May 2021, doi:
  \href{https://doi.org/10.1109/TNS.2021.3065122}{10.1109/TNS.2021.3065122}.

\bibitem{martinella2023}
C.~Martinella, S.~Race, R.~Stark, R.~G. Alia, A.~Javanainen, and U.~Grossner,
  ``High-energy proton and atmospheric-neutron irradiations of {S}i{C} power
  {MOSFET}s: {SEB} study and impact on channel and drift resistances,''
  \emph{IEEE Trans. Nucl. Sci.}, vol.~70, no.~8, pp. 1844--1851, Aug. 2023,
  doi:
  \href{https://doi.org/10.1109/TNS.2023.3267144}{10.1109/TNS.2023.3267144}.

\bibitem{alberton2025dtneutrons}
S.~G. Alberton, A.~C.~V. Bôas, M.~A. Guazzelli, V.~A.~P. Aguiar, N.~Added,
  C.~A. Federico, T.~C. Cavalcante, E.~C.~F. Pereira~Júnior, R.~G. Vaz, O.~L.
  Gonçalez, J.~Wyss, A.~Paccagnella, and N.~Medina, ``Single-event effects
  induced by monoenergetic fast neutrons in silicon power {UMOSFET}s,''
  submitted for publication. \textit{TechRxiv}. December 23, 2024., doi:
  \href{https://doi.org/10.36227/techrxiv.173496316.61556618/v1}{10.36227/techrxiv.173496316.61556618/v1}.

\bibitem{alberton2022radecs}
S.~G. Alberton, A.~C.~V. Bôas, N.~H. Medina, M.~A. Guazzelli, V.~A.~P. Aguiar,
  N.~Added, C.~A. Federico, O.~L. Gonçalez, T.~C. Cavalcante, E.~C.~F.
  Pereira~Júnior, and R.~G. Vaz, ``Alpha particle- and neutron-induced
  single-event effects in {COTS} power {FET}s,'' in \emph{2022 22nd European
  Conference on Radiation and Its Effects on Components and Systems (RADECS)},
  2022, pp. 1--4, doi:
  \href{https://doi.org/10.1109/RADECS55911.2022.10412582}{10.1109/RADECS55911.2022.10412582}.

\bibitem{alberton2024radecs}
S.~G. Alberton, A.~C.~V. Bôas, M.~A. Guazzelli, J.~Wyss, V.~A.~P. Aguiar,
  N.~Added, C.~A. Federico, T.~C. Cavalcante, E.~C.~F. Pereira~Júnior, R.~G.
  Vaz, O.~L. Gonçalez, L.~Silvestrin, S.~Mattiazzo, C.~Cazzaniga,
  M.~Kastriotou, C.~Frost, A.~Paccagnella, and N.~Medina, ``Studies on the
  reliability of power {UMOSFET}s in the terrestrial radiation environment,''
  to be published. \textit{2024 24nd European Conference on Radiation and Its
  Effects on Components and Systems (RADECS)}.

\bibitem{cazzaniga2018}
C.~Cazzaniga, M.~Bagatin, S.~Gerardin, A.~Costantino, and C.~D. Frost, ``First
  tests of a new facility for device-level, board-level and system-level
  neutron irradiation of microelectronics,'' \emph{IEEE Trans. Emerg. Topics
  Comput.}, vol.~9, no.~1, pp. 104--108, Jan.-Mar. 2021, doi:
  \href{https://doi.org/10.1109/TETC.2018.2879027}{10.1109/TETC.2018.2879027}.

\bibitem{chiesa2018}
D.~Chiesa, M.~Nastasi, C.~Cazzaniga, M.~Rebai, L.~Arcidiacono, E.~Previtali,
  G.~Gorini, and C.~D. Frost, ``Measurement of the neutron flux at spallation
  sources using multi-foil activation,'' \emph{Nucl. Instrum. Methods Phys.
  Res. A}, vol. 902, pp. 14--24, Sep. 2018, doi:
  \href{https://doi.org/10.1016/j.nima.2018.06.016}{10.1016/j.nima.2018.06.016}.

\bibitem{jedec-jesd89b}
{JEDEC-JESD89B}, ``Measurement and reporting of alpha particle and terrestrial
  cosmic ray induced soft errors in semiconductor devices,'' JEDEC Solid State
  Technology Association, Tech. Rep., 2021, [Online]. Available:
  \href{https://www.jedec.org/standards-documents/docs/jesd-89a}{https://www.jedec.org/standards-documents/docs/jesd-89a}.

\bibitem{gordon2004}
M.~Gordon, P.~Goldhagen, K.~Rodbell, T.~Zabel, H.~Tang, J.~Clem, and P.~Bailey,
  ``Measurement of the flux and energy spectrum of cosmic-ray induced neutrons
  on the ground,'' \emph{IEEE Trans. Nucl. Sci.}, vol.~51, no.~6, pp.
  3427--3434, Dec. 2004, doi:
  \href{https://doi.org/10.1109/TNS.2004.839134}{10.1109/TNS.2004.839134}.

\bibitem{lucsanyi2022}
D.~Lucs{\'a}nyi, R.~G. Al{\'\i}a, K.~Bi{\l}ko, M.~Cecchetto, S.~Fiore, and
  E.~Pirovano, ``{G4SEE}: {A} {G}eant4-based single event effect simulation
  toolkit and its validation through monoenergetic neutron measurements,''
  \emph{IEEE Transactions on Nuclear Science}, vol.~69, no.~3, pp. 273--281,
  Mar. 2022, doi:
  \href{https://doi.org/10.1109/TNS.2022.3149989}{10.1109/TNS.2022.3149989}.

\bibitem{michez2015}
A.~Michez, S.~Dhombres, and J.~Boch, ``{ECORCE}: {A} {TCAD} tool for total
  ionizing dose and single event effect modeling,'' \emph{IEEE Trans. Nucl.
  Sci.}, vol.~62, no.~4, pp. 1516--1527, Aug. 2015, doi:
  \href{https://doi.org/10.1109/TNS.2015.2449281}{10.1109/TNS.2015.2449281}.

\bibitem{baliga2018}
B.~J. Baliga, \emph{Fundamentals of Power Semiconductor Devices}, 2nd~ed.\hskip
  1em plus 0.5em minus 0.4em\relax Springer International Publishing, 2018.

\bibitem{baliga2010}
------, \emph{Advanced {P}ower {MOSFET} {C}oncepts}.\hskip 1em plus 0.5em minus
  0.4em\relax Springer Science \& Business Media, 2010.

\bibitem{milstd750e}
{Department of Defense (USA)}, ``Test method standard: {T}est methods for
  semiconductor devices ({MIL}-{STD}-750{E}),'' Tech. Rep., 2006, [Online].
  Available: \href{https://quicksearch.dla.mil}{https://quicksearch.dla.mil}.

\bibitem{liu2012}
S.~Liu, R.~Marec, P.~Sherman, J.~L. Titus, F.~Bezerra, V.~Ferlet-Cavois,
  M.~Marin, N.~Sukhaseum, F.~Widmer, M.~Muschitiello \emph{et~al.},
  ``Evaluation on protective single event burnout test method for power
  {DMOSFET}s,'' \emph{IEEE Trans. Nucl. Sci.}, vol.~59, no.~4, pp. 1125--1129,
  2012, doi:
  \href{https://doi.org/10.1109/TNS.2011.2177863}{10.1109/TNS.2011.2177863}.

\bibitem{lee2011}
J.~C. Lee and N.~J. McCormick, \emph{Risk and Safety Analysis of Nuclear
  Systems}.\hskip 1em plus 0.5em minus 0.4em\relax John Wiley \& Sons, 2011.

\bibitem{alberton2022}
S.~G. Alberton, V.~A.~P. Aguiar, N.~H. Medina, N.~Added, E.~L.~A. Macchione,
  R.~Menegasso, G.~J. Ces{\'a}rio \emph{et~al.}, ``Charge deposition analysis
  of heavy-ion-induced single-event burnout in low-voltage power {VDMOSFET},''
  \emph{Microelectron. Reliab.}, vol. 137, no. 114784, pp. 1--7, Oct. 2022,
  doi:
  \href{https://doi.org/10.1016/j.microrel.2022.114784}{10.1016/j.microrel.2022.114784}.

\bibitem{sahu2003}
K.~Sahu, ``{EEE-INST-002}: {I}nstructions for {EEE} parts selection, screening,
  qualification, and derating,'' NASA/TP-2003-212242, Tech. Rep., 2003,
  [Online]. Available:
  \href{https://nepp.nasa.gov/pages/EEE-INST-002.cfm}{https://nepp.nasa.gov/pages/EEE-INST-002.cfm}.

\bibitem{jedec-jep151a}
{JEDEC-JEP151A}, ``Test procedure for the measurement of terrestrial cosmic ray
  induced destructive effects in power semiconductor devices,'' JEDEC Solid
  State Technology Association, Tech. Rep., 2022, [Online]. Available:
  \href{https://www.jedec.org/standards-documents/docs/jep151}{https://www.jedec.org/standards-documents/docs/jep151}.

\bibitem{kuboyama2004}
S.~Kuboyama, N.~Ikeda, T.~Hirao, and S.~Matsuda, ``Improved model for
  single-event burnout mechanism,'' \emph{IEEE Trans. Nucl. Sci.}, vol.~51,
  no.~6, pp. 3336--3341, Dec. 2004, doi:
  \href{https://doi.org/10.1109/TNS.2004.839512}{10.1109/TNS.2004.839512}.

\bibitem{egawa1966}
H.~Egawa, ``Avalanche characteristics and failure mechanism of high voltage
  diodes,'' \emph{IEEE Trans. Electron Devices}, vol. ED-13, no.~11, pp.
  754--758, Nov. 1966, doi:
  \href{https://doi.org/10.1109/T-ED.1966.15838}{10.1109/T-ED.1966.15838}.

\bibitem{wrobel1985}
T.~F. Wrobel, F.~N. Coppage, G.~L. Hash, and A.~J. Smith, ``Current induced
  avalanche in epitaxial structures,'' \emph{IEEE Trans. Nucl. Sci.}, vol.~32,
  no.~6, pp. 3991--3995, Dec. 1985, doi:
  \href{https://doi.org/10.1109/TNS.1985.4334056}{10.1109/TNS.1985.4334056}.

\bibitem{tambone2024}
R.~Tambone, A.~Ferrara, R.~Siemieniec, A.~Wood, F.~Magrini, and R.~J.~E.
  Hueting, ``Ruggedness of silicon power {MOSFET}s—{P}art {I}: {C}ell
  structure design related failure: {A} review,'' \emph{IEEE Trans. Electron
  Devices}, vol.~71, no.~6, pp. 3445--3457, Jun. 2024, doi:
  \href{https://doi.org/10.1109/TED.2024.3394452}{10.1109/TED.2024.3394452}.

\bibitem{liu2007}
S.~Liu, J.~L. Titus, and M.~Boden, ``Effect of buffer layer on single-event
  burnout of power {DMOSFET}s,'' \emph{IEEE Trans. Nucl. Sci.}, vol.~54, no.~6,
  pp. 2554--2560, Dec. 2007, doi:
  \href{https://doi.org/10.1109/TNS.2007.910869}{10.1109/TNS.2007.910869}.

\bibitem{liu2006}
S.~Liu, M.~Boden, D.~A. Girdhar, and J.~L. Titus, ``Single-event burnout and
  avalanche characteristics of power {DMOSFET}s,'' \emph{IEEE Trans. Nucl.
  Sci.}, vol.~53, no.~6, pp. 3379--3385, Dec. 2006, doi:
  \href{https://doi.org/10.1109/TNS.2006.884971}{10.1109/TNS.2006.884971}.

\bibitem{ferlet2010}
V.~Ferlet-Cavrois, F.~Sturesson, A.~Zadeh, G.~Santin, P.~Truscott, C.~Poivey,
  J.~R. Schwank, D.~Peyre, C.~Binois, T.~Beutier, A.~Luu, M.~Poizat,
  G.~Chaumont, R.~Harboe-S?rensen, F.~Bezerra, and R.~Ecoffet, ``Charge
  collection in power {MOSFET}s for {SEB} characterisation—{E}vidence of
  energy effects,'' \emph{IEEE Trans. Nucl. Sci.}, vol.~57, no.~6, pp.
  3515--3527, 2010, doi:
  \href{https://doi.org/10.1109/TNS.2010.2086077}{10.1109/TNS.2010.2086077}.

\bibitem{grasser2014}
T.~Grasser, \emph{Hot Carrier Degradation in Semiconductor Devices}.\hskip 1em
  plus 0.5em minus 0.4em\relax Springer, 2014.

\bibitem{sze2021}
S.~M. Sze, Y.~Li, and K.~K. Ng, \emph{Physics of Semiconductor Devices},
  4th~ed.\hskip 1em plus 0.5em minus 0.4em\relax John Wiley \& Sons, 2021.

\end{thebibliography}

\end{document}